# Magnetic tubules

Damien Letellier[1,*], Olivier Sandre[2], Christine Ménager[1], Valérie Cabuil[1], Michel Lavergne[3]

[1] Laboratoire des Liquides Ioniques et Interfaces Chargées, Université Paris 6, Bât F, Case 63, 4 Place Jussieu, 75252 Paris cedex 05 – France
[2] Laboratoire des Matériaux Désordonnés et Hérérogènes, Université Paris 6, URA CNRS, Tour 13, Case 78, 75252 Paris cedex 05 – France
[3] Service de Microscopie Electronique, Groupement Régional de Mesures Physiques, Université Paris 6, Bât F, 75252 Paris cedex 05 – France

* corresponding author

*Abstract*

Dispersion of tubules made of diacetylenic phospholipids ($DC_{8,9}PC$), in aqueous colloidal dispersions of magnetic nanoparticles is studied as a function of the sign of particles surface charges. In every case the tubule-vesicle transition temperature is decreased by the presence of the magnetic nanoparticles. Electrophoresis experiments on the tubules in pure water permits to conclude on a negative apparent surface charge. We study the magnetic response of the system to a static or stationary rotating field and to a magnetic field gradient. These experiments reveal an excess or a lack of magnetic permeability between tubules and the surrounding medium. Electron microscopy confirms these results showing an electrostatic interaction between the phospholipidic bilayer and the magnetic particles.

## Introduction

The present work describes the use of molecular selfassemblies of amphiphilic molecules to synthesize new microstructures which exhibit original properties suitable for technical applications. Many recent works report the use of molecular self assemblies of lipids molecules as templates for the crystallization of inorganic oxides in order to obtain new morphologies [1,2] or to synthesize porous materials [3,4]. Due to their morphology, size, hollowness and rigidity, lipidic tubules composed of multibilayers wrapped in a cylinder [5] are well suited for technical applications [6]. In the particular case of magnetic applications, anisotropy and rigidity are both required and an electroless deposition technique has already been described in order to obtain permalloy-coated tubules [7]. Here we describe the synthesis of tubules which get a magnetic behavior when they are well combined with magnetic colloids made of nanoscopic magnetic particles suspended in a liquid solvent [8].

The compatibility between self-assemblies of tensioactive molecules with magnetic colloidal dispersions has already been considered. The first hybrid system combining a lyotropic lamellar phase and magnetic particles was described by Fabre et al. [9]. In this system the tensioactive membrane was fluid. as the one of the liposomes into which we had then introduced magnetic particles [10]. The main problem for the synthesis of the so-called magnetic liposomes was the impossibility to encapsulate high volume fractions of particles because liposomes can not afford the heavy osmotic pressure due to the high ionic strength of the encapsulated magnetic dispersions. In the case of magnetic tubules the rigidity of the membrane avoids this problem. Using magnetic fluids with a high volume fraction of particles allows to observe a response of tubules to low intensity magnetic fields. We interpret the magnetic behavior of the magnetized tubules as monitored by the nature of the interaction between the phospholipidic membrane and the magnetic nanoparticles coated either by positive or negative surface charges.

## 1. Materials and methods

*1.1 The tubules*

Observation of tubules has been described in systems containing synthetic chiral amphiphiles [11], amino-acid-based surfactants [12], sugar-based surfactants [13] and biological galactocerebrosides [14]. Tubules formation is driven by a reversible first order phase transition from an interlamellar chain-melted $L_\alpha$ phase to a chain-frozen $L_{\beta'}$ phase, the tubule phase being the one in which the hydrocarbon chains are highly ordered. Here we use a symmetric phosphatidylcholine lipid: 1,2-bis(10,12-tricosadiynoyl)-*sn*-glycero-3-phosphocholine or $DC_{8,9}PC$ obtained from Avanti Polar Lipids and used as purchased. Polymerizable lipids with diacetylenic fatty acyl chains as $DC_{8,9}PC$ are known to form tubular microstructures when Iiposornes of these lipids are cooled through their chain melting transition [11]. $DC_{8,9}PC$ chain melting transition temperature is approximately 43°C [5]. Above this temperature $DC_{8,9}PC$ forms liposomes in aqueous dispersion. To obtain tubules we employ the precipitation method firstly described by Schoen and co-workers [5] which consists in the solubilisation of lipids in an appropriate solvent and a further precipitation by addition of water.

The method is as follows: a solution of lipids $DC_{8,9}PC$ (1 mg) is dissolved in ethanol (1 mL) at room temperature. Then 1 mL of water is added drop by drop, a flocculent precipitate of tubules is observed. This precipitate is centrifuged at 6000 rpm for 15 min and the resulting pellet is dispersed in distilled water.

*1.2 The magnetic fluids*

An aqueous magnetic fluid is a colloidal dispersion of magnetic ferric oxide particles in water [15]. Magnetite particles are firstly synthesized according to Massart' s procedure [16], then oxidized to rnaghernite by ferric nitrate in nitric acid medium. Particles mean diameter is about 7 nm. These particles have surface charges due to the acid-base behavior of their surface. According to the pH, the sign and the density of surface charges can be monitored.

We used here two kinds of particles: i) cationic particles, which have a positive surface charge density around $0.2$ C m$^{-2}$ in acidic media (pH~2); ii) anionic particles, which are coated by citrate species and have negative charges (around $0.2$ C m$^{-2}$) for pH~7.

These nanometric particles dispersed in water lead to aqueous colloidal dispersions (called ionic ferrofluid) stabilized by the electrostatic repulsions between the magnetic grains. No additional surfactant is needed. Particles never separate from the solvent, even when submitted to a high magnetic field gradient.

*1.3. Mixtures of tubules with magnetic fluid*

Aqueous magnetic fluids can replace the normal aqueous phase in organized systems of surfactant molecules provided that their ionic strength does not destabilize the lyotropic system [17]. That is the reason why we have dialyzed all our magnetic dispersions through a membrane of Visking type until their conductivity is inferior to 2 mS cm$^{-1}$. Then DCPC tubules have been precipitated in water as described above and dispersed, after centrifugation, in the dialyzed magnetic dispersions.

*1.4. Observation of tubules behavior*

All the methods we shall describe further are grounded on observation of tubules under an optical microscope. The sample is kept in a sealed glass capillary with a rectangular crosssection (50 μm × 1 mm or 200 μm × 4 mm). According to the geometry of the experiment, a normal (Zeiss, Ortholuxe) or an inverted microscope (Leica, LEITZ DMIL) is used.

The difference between refraction indices of DCPC tubules and solvent (pure water or magnetic fluid) aIlows image formation without the use of phase contrast. When it is possible to approach the objective near the sample, a magnification ×40 is used, with a numerical aperture (NA) of 0.75 or 0.65. Otherwise a magnification ×32 with NA equal to 0.40 or 0.30 is sufficient. The tubules are filmed with a charge coupled device (CCD) color camera (Vista, VPC 4130, U.K.) connected to a video tape recorder. Selected images are digitized on a computer (768 × 512 pixels, 256 gray levels) and processed with the NIH Image (National Institute of Health, USA) software. Noise reduction of the pictures is obtained with a standard algorithm (median filter).

*1.4.1. Electrophoresis*

The simple setup we used is schematized on Fig. I. Two copper wires are immersed in a glass capillary, the distance between the electrodes being approximately 4mm. The applied potential is a few Volts. In spite of its simplicity, this basic method gives qualitative informations on the electrokinetic behavior of the tubules. Nevertheless the drawbacks are the limitation of the applied voltage by water electrolysis and the lack of control of some parameters: geometry of the current lines, electroosmosis caused by the negative charge of the glass walls. That is why the results are controIled with a more sophisticated experiment built by L. Mitnik et al. [18,19]. In their setup two solvent reservoirs at different electric potentials are tightly isolated, except by a bridge in which the current lines are confined. This one is made of two glass slides separated by a spacer, containing the solution to be studied . Solvent electrolysis is no longer a problem because it appears next to the electrodes which are far from the region of observation. Tensions as high as 100 V can be applied, provided that the conductivity is low enough (18 MΩ em resistivity water). Electroosmosis (spontaneous motion of the solvent in a capillary under an electric field just because the waIls bear a surface charge) is a phenomenon which is more difficult to be independent of. Two strategies have been tested: coating the glass surfaces with bovine serum albumin (BSA), or closing the cell with an agarose plug at each end [19].

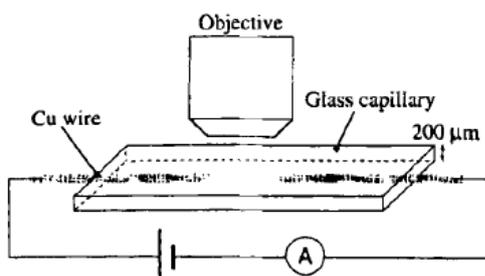

Fig. 1. Electrophoresis basic experiment. The electric field is about 10$^3$ V m$^{-1}$.

*1.4.2. Magnetic alignment: fixed or rotating magnetic field*

The glass capillary containing tubules in a magnetic fluid is set between two identical coils in serial (125 turns each, self-inductance equals to 2.5 mH). With polar pieces (soft iron) inside the coils, the magnetic field can be varied from $2 \times 10^3$ to $6.4 \times 10^4$ A m$^{-1}$ (25 to 800 Gauss), and for each intensity the mean time of alignment is measured.

Another experiment to test the orienting of tubules consists in applying a rotating magnetic field with pulsation Ω. In that experiment the sample is in a square cell made of two glass coverslides (20mm side) spaced by a polymer film based on paraffin (Parafilm) after melting on a heating plate. The field is created by two pairs of coils oriented at 90° (Fig. 2). Circular polarisation is obtained when the currents $I_X$ and $I_y$ through the coils are in phase quadrature. The phase of the rotating magnetic field is known thanks to a periodic light pulse (time width: 6ms) superposed to the image in the microscope and triggered with $I_X$.

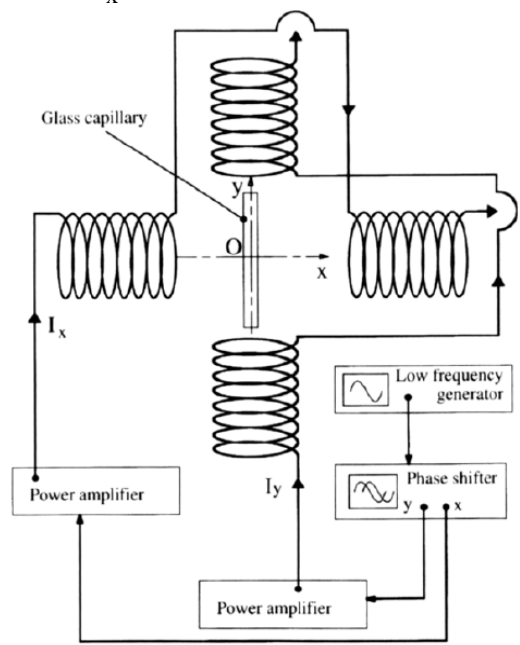

Fig. 2. Magnetic alignment by a rotating magnetic field. The two pairs of coils and the appropriate electronic setup are sketched.

*1.4.3 Magnetophoresis*

The aim is to observe a migration under a magnetic field gradient. The setup is described on Fig. 3. A magnetic field is applied along Ox-direction and aligns tubules parallel to Ox. A gradient is created perpendicular to the field direction (Oy) by deviating the field lines of two coils (125 turns each, self-inductance equal to 2.5mH) in serial with polar pieces edged at 45°. Measurements with Hall effect probes give a gradient of $9.6 \times 10^5$ A m$^{-2}$ (120 G/cm) with a mean field equal to $2.4 \times 10^3$ A m$^{-1}$ (300G) under a 3A direct current.

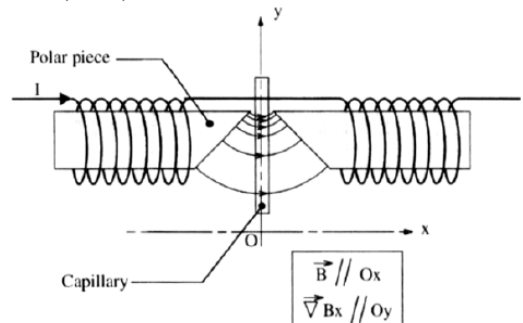

Fig. 3. Magnetophoresis experiment. The field lines are parallel one to each other in the region of the capillary and the field gradient is perpendicular to them.

*1.4.4. Thermolysis of tubules*

Thermoregulation of sealed cappilaries is obtained using a usual Peltier cell with a platinum probe and a water cooling. Less common, a heating by induction was build. A radio-frequency magnetic field is produced with coils in an electric resonant circuit. The resonance occurs at 850 kHz. A cooling circuit with nonane is used instead of water because its high dielectric constant hugely decreases the quality factor.

*1.4.5. Transmission electron microscopy (TEM)*

TEM is performed with a JEOL 100CXII top entry UHR microscope, on samples deposited on carbon films and air dried. Visualization of the helical structure of the tubules is enhanced when samples are negatively stained with ammonium molybdate according to the procedure described in ref [20]. Nevertheless, tubules combined with magnetic particles have also been directly observed without staining, for magnetic particles provide enough contrast to make the tubules observable.

## 2. Results

*2.1. Characteristic of DCPC tubules in pure water*

We have firstly characterized the tubules obtained as described in section A in terms of length. thickness and surface charge.

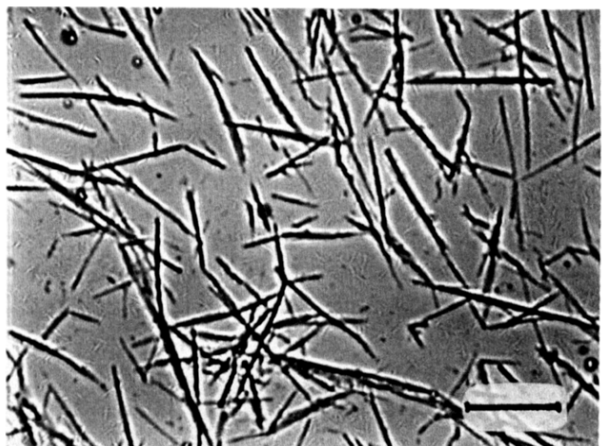

Fig. 4. Optical micrograph picture of the tubules which appear as black cylinders with a relatively broad distribution of lengths. The bar represents 20 μm.

Optical microscopy (Fig. 4) reveals that tubules are long with a relatively broad distribution of lengths (few microns to few tenths of microns). It can be noticed that the synthesis we used does not allow to control precisely the length of tubules, but this parameter could be controlled according to Ref. [21] in which the autors describe the synthesis of tubules formed under controlled cooling rates.

On the other hand, TEM shows that tubules' diameter is constant throughout all the samples, with an average value of 0.5 μm. Tubules appear like hollow cylinders which can be made of one or much bilayers. We observed tubules and helices which appear to have only one bilayer per wall (Fig.5). Nevertheless such tubules consisting of a single cylindrical bilayer are as straight and rigid as those with multiple bilayer walls.

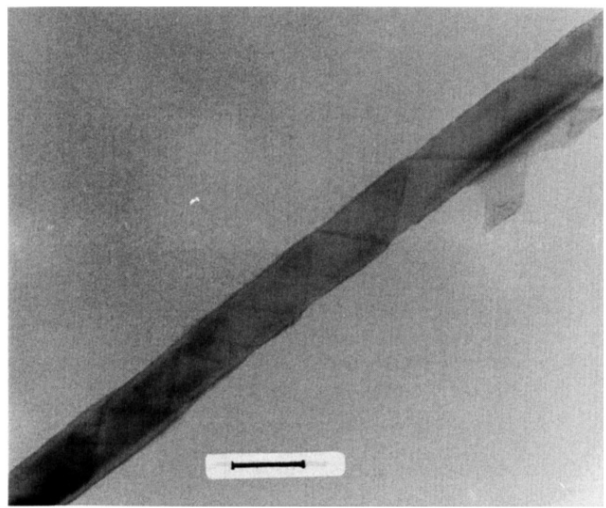

Fig. 5. Electron micrograph image (negative stain microscopy) of a $DC_{8,9}PC$ tubule. This one seems to have one bilayer wall only. This bilayer is rolled so as to delimit a hollow cylinder. The bar represents 1 μm.

When an electric field is applied on DCPC tubules in pure water, they move towards the positive electrode. For a field intensity of the order $10^3$ V m$^{-1}$ (4 V between the eletrodes separated by 4 mm), their velocity is approximately 30 μm s$^{-1}$. Fig. 6a gives a schematic velocity profile: almost all the tubules move in the opposite direction of the applied field, except the ones which adhere to the glass walls (they are still) and those which are in a thin region next to the walls (these latter move towards the negative electrode). This backflow is much smaller than the principal flow, so that the net current is towards the anode. The dependency of this electocinetik behavior with the Debye lenght $\kappa^{-1}$ is tested by using tubules prepared in ultrapure water (18 MΩ cm resistivity obtained thanks to Millipore filters: $\kappa^{-1}$ up to 1 mm) instead of simply distillated water. While the measured electric current is unchanged (I=2 mA for a 4 V voltage), the tubules migration is affected: the direction of the net motion is unclear, and some trajectories of tubules make loops. The experiment on Mitnik's setup gives the same kind of velocity profile in the case when the capillary is end up by agarose plugs. When the glass slides are covered by a layer of BSA proteins, the only difference is that the backflow near the glass walls is not observed anymore.

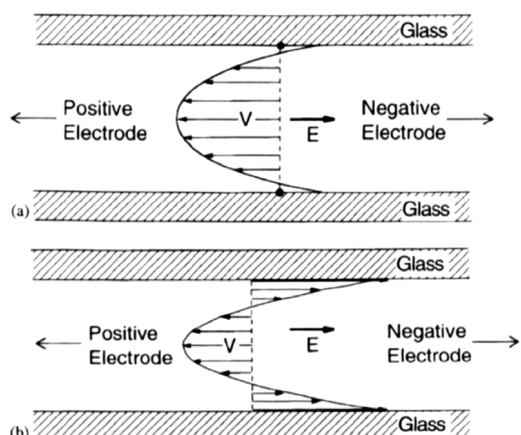

Fig. 6. (a) Schematic experimental velocity profile of tubules in pure water under an applied electric field. The surface backflow is much smaller than the principal flow. (b) Electroosmosis velocity profile with closed capillary conditions as described in reference [19]. The profile is parabolic and the total flow is zero.

## 2.2. Tubules dispersed with anionic magnetic particles in water

Tubules are mixed with a dialyzed magnetic fluid at pH ~ 7 as described in the experimental section. The addition of the magnetic fluid does not cause any damage to the tubules according both to optical and electronic microscopy. TEM observations are performed without the use of the coloring agent; as a matter of fact ammonium, molybdate induces precipitation of the particles on the tubules surface (Fig. 7). Without coloration, the nanoparticles appear homogeneously dispersed in the solution: they form a uniform film that covers more or less the membrane of the tubule (Fig. 8).

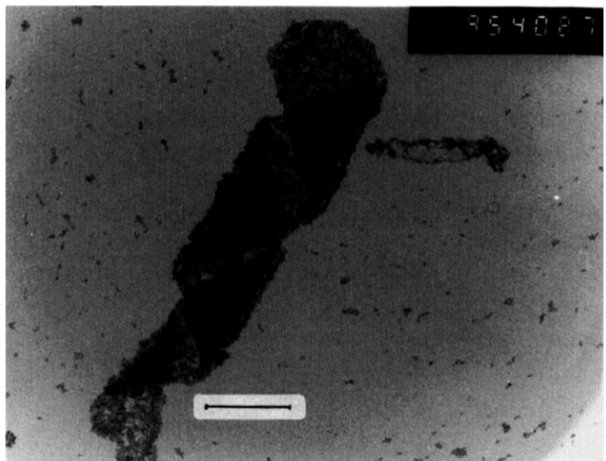

Fig. 7. Negative stain electron micrograph of a helix (unrolled tubule) dispersed in an anionic magnetic fluid. The coloring agent induces the precipitat ion of the particles on the membrane. The bar represents 0.5 µm.

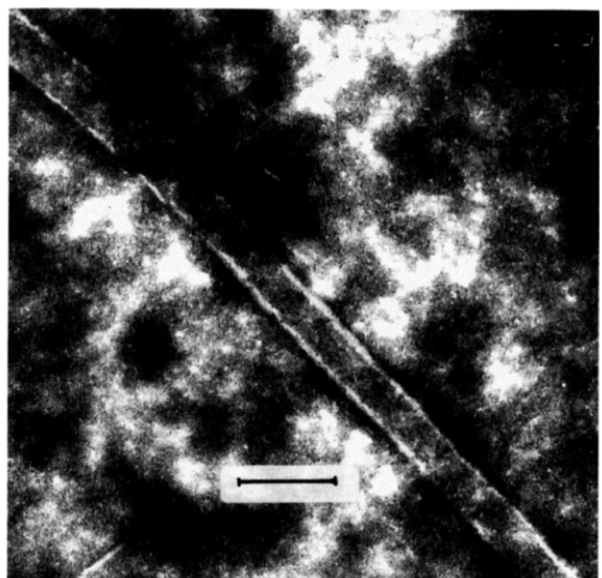

Fig. 8. Electron micrograph image, without coloration, of a tubule combined with the same anionic magnetic fluid as in Fig. 7. The particles form a uniform film which covers partially the membrane. Bar length is 1 µm.

When a magnetic field is applied to the mixture, tubules align themselves along the field lines (Fig. 9). Orientation times have been measured and depend on the volume fraction of magnetic particles and on the field intensity. Below a critical value of the magnetic field intensity or of the particle's concentration, orientation does not occur because the magnetic torque is too weak compared to thermal agitation. Typical times for alignment are given in Fig. 10a.

In the magnetophoresis experiment, the DCPC tubules are oriented in the field direction (Ox) and migrate along the direction of the field gradient (Oy). When dispersed in an anionic magnetic fluid (constituted by negative particles such as the volume fraction of particles is around 1.8 %), the motion of tubules is always in the direction of decreasing field intensity (towards negative $y$ values, Fig. 11). A typical value of velocity is 1.5 µm s$^{-1}$ and it increases with the field gradient.

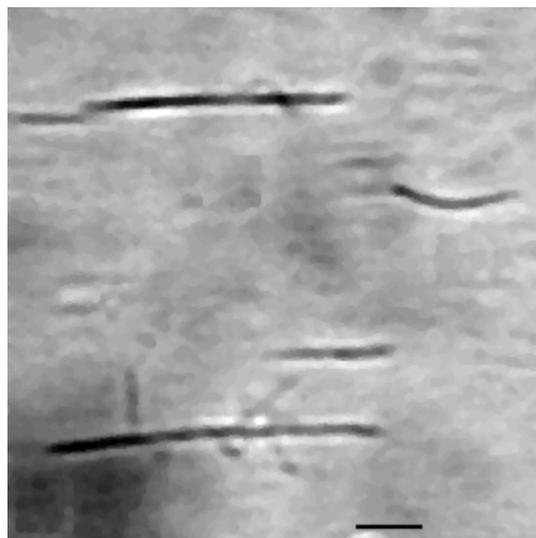

Fig. 9. Optical micrograph of tubules alignment under a field of 2.4 × 10$^3$ A m$^{-1}$ (300G). Bar length is 5 µm.

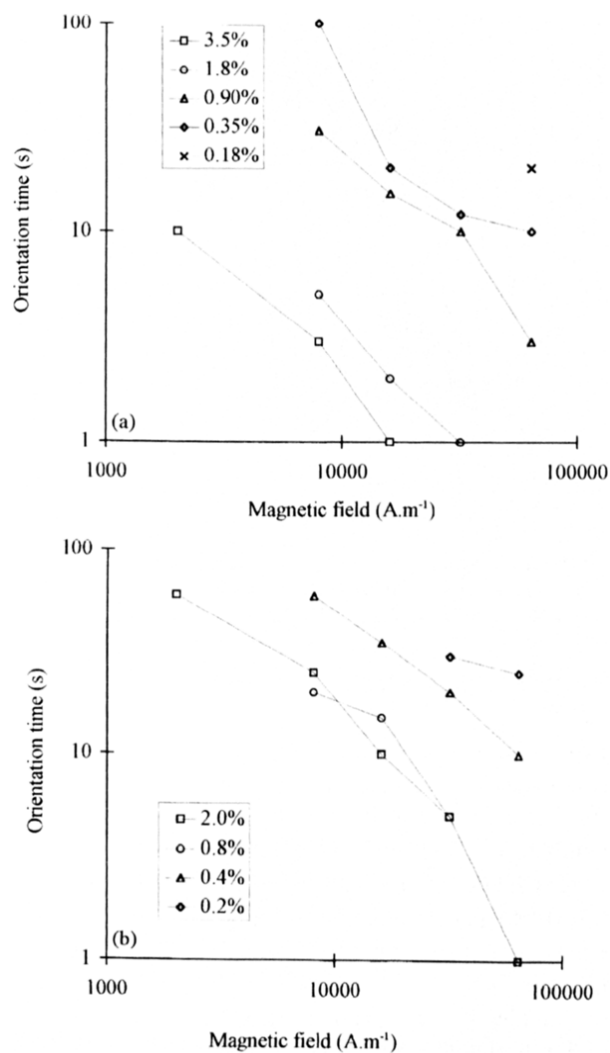

Fig. 10. ( a ) Orientation time of tubules in an anionic magnet ic fluid as a function of the applied magnetic field H. The symbols correspond to various volume fractions ϕ in magnetic particles. (b) Orientation time of tubules in a cationic magnetic fluid as a function of the applied magnetic field H. The symbols correspond to various volume fractions ϕ in magnetic particles. The fields allowing to align the tubules are higher than for Fig. l0a.

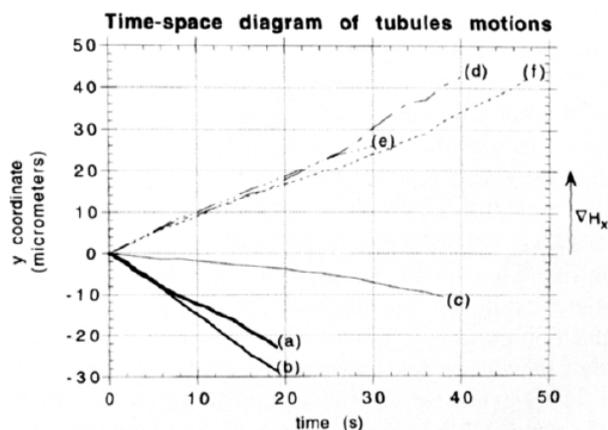

Fig. 11. Time-space diagram of tubules magnetophoretic motion s: (a) (b) (c) tubules in an anionic magnetic fluid ($\phi = 1.8$ %) at several field gradient values (a = 9.6, b = 6.6, and c= $3.1 \times 10^5$ A m$^{-2}$) ; (d) (e) (f) tubule, in a cationic magnetic fluid ($\phi = 2$ %) at a given field gradient ($9.6 \times 10^5$ A m$^{-2}$) for several tubule lenghts (d = 9, e = 21 and f = 27 μm).

Submitted to a magnetic field with circular polarization, the tubules rotate synchronously with the field up to a critical angular velocity $\Omega_C$ (Fig. 12). For a volume fraction of particles equal to 3.5 % and a field intensity of $4.0 \times 10^3$ A m$^{-1}$ (51 G) $\Omega_C/2\pi$ is 0.5Hz ($\Omega_C$=3.14 rad s$^{-1}$). Fig. 12 shows a tubule rotating at 0.1Hz: the angle increment every quarter of field period (T/4= 2.5 s) being always 90°, the rotation is perfectly synchronous. However, at field frequencies higher than $\Omega_C/2\pi$, the rotation of the tubules is not synchronous anymore, the coupling to the field rotation being only intermittent (motion described as jingle movement by Skjeltorp et al. [22] ). It can be noticed that the critical rotation period is 2s, a value which is not so far from the time of alignment (3 s) of a tubule in the same magnetic fluid under 100G.

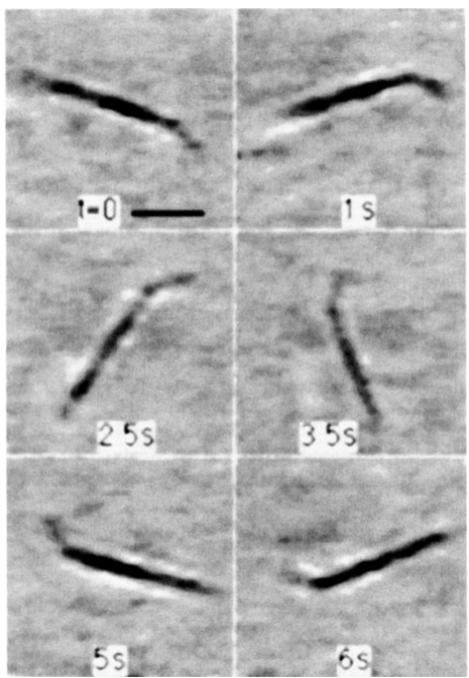

Fig. 12. Synchronous rotation of a tubule in an anionic magnetic fluid ($\phi = 3.5$%) under a $6.3 \times 10^3$ A m$^{-1}$ (80 G) field rotating at 0.1 Hz. Theperiod T is l0 s and the time increment between two consecutive pictures of the same column is T/4. The bar length is 5 μm.

Finally, it appeared that the thermal behavior ofthe DCPC bilayers is strongly modified by the presence of anionic particles. The tubule-vesicle transition temperature is lowered in a wide range: Tm ~ 10°C lower in a magnetic fluid (volume fraction in particles 3.5%) than in pure water (Fig. 13).

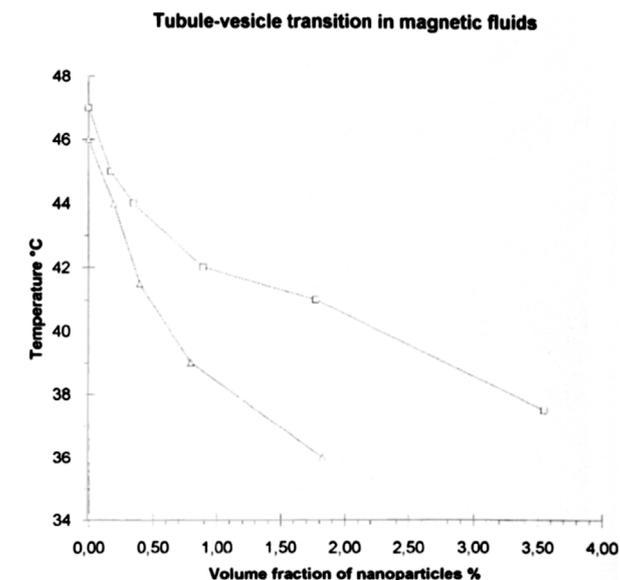

a

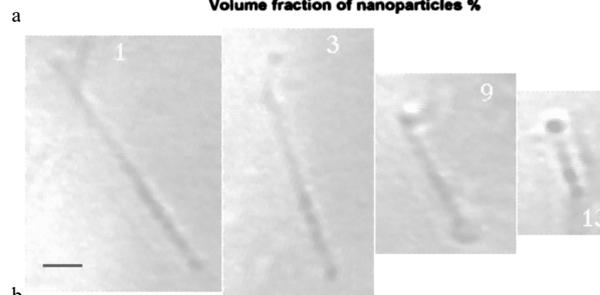

b

Fig. 13. (a) Tubule-vesicle transition temperature *versus* volume fraction $\phi$ of magnetic nanoparticles in the case of an anionic magnetic fluid (squares) and in the case of a cationic magnetic fluid (triangles). (b) Images of a tubule undergoing transition under optical microscopy.

### 2.3 Tubules dispersed with cationic magnetic particles in water

On the microscopic scale, the tubules dispersed in a cationic magnetic fluid have the same appearance and the same behavior as tubules dispersed in an anionic one. However electron microscopy (without coloration) shows that in this case, the tubules appear like black cylinders made of concentrated magnetic particles (Fig . 14a) in a diluted dispersion of particles. It can be noticed that the concentrations of the preparation for picture 8 and 14a are of the same order. At a higher magnification (Fig. 14b), one can see the lipid membrane which appears like two white bands delimit ing the hollow tube full of particles. At the extremity (Fig. 14c), the tubules seem to bring out the particles.

In the magnetophoresis experiment, tubules with cationic particles exhibit an opposite behavior compared to the tubules with anionic particles: almost all of them migrate in the direction of increasing field intensity (towards positive *y* values, Fig. 11). Nevertheless a small amount of the tubules still goes in the opposite direction. The velocities under a $9.6 \times 10^5$ A m$^{-2}$ gradient range from 0.3 to 0.9 μm s$^{-1}$.

As in the former case, tubules in the cationic magnetic fluid ($\phi = 2$ %, pH=3.55) can rotate synchronously with the magnetic field (Fig. 15), but the critical frequency is ten times smaller: 0.05Hz ($\Omega_C$=0.314 rad s$^{-1}$) under a field of $6.3 \times 10^3$ A m$^{-1}$ (80 G). The phase lag between the tubule and the field is 5° at 0.01Hz and 38° at 0.05Hz, the maximum phase lag been 45° (see discussion below). Once again the minimal rotation period (20s) is approximately the time of alignment of a tubule (25 s) measured at $8 \times 10^3$ A m$^{-1}$ (100G) (Fig. 10b).

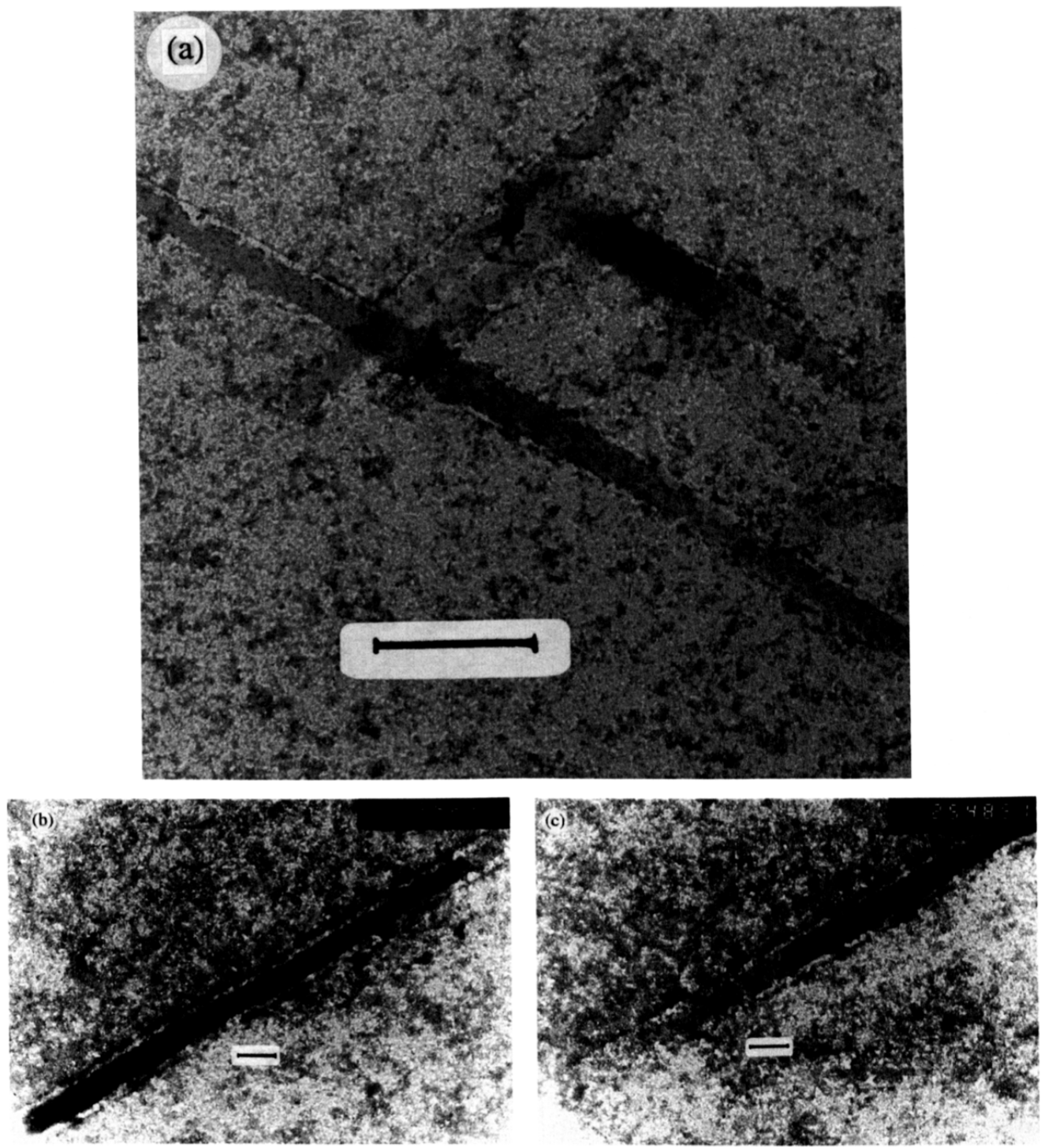

Fig. 14. Electron micrograph pictures of tubules with cationic magnetic particles for different magnifications: a) Tubules appear like black cylinders full of particles in a diluted dispersion of the same particles. Bar length is 2 µm. b) This magnification permits to observe the membrane (two white bands) delimiting the walls of the cylinder containing the particles. Bar length is 0.5 µm. c) At the extremity the tubule seems to bring out the panicles. Bar length is 0.5 µm.

The transition temperature of the tubules in an acidic magnetic fluid is also affected by the presence of the cationic particles. This transition leads to globular aggregates badly defined under microscope observation (Fig. 13b). In the case of cationic particles as with anionic particles, TEM observations of heated samples exhibit some vesicles.

## 3. Discussion

The experimental results described above give information concerning DCPC tubules and their behavior when they are associated to magnetic nanoparticles. The combining of the different experiments allows to get some conclusions.

### 3.1. Charges on DCPC tubules in water

The question of tubules' charge is very important for they will be mixed with other charged species. DCPC is a zwitterionic molecule thus the structural charge of tubules should be zero. In the electrophoresis experiment it may be discussed whether the tubules have an own electrokinetic behavior or if the motion observed is an artifact due to electroosmosis. In the electroosmosis profile between two planar walls with closed capillary condition s (Fig. 6b), the surface velocity (towards cathode) is two times higher than the backflow velocity (towards anode) in the middle of the cell [18, 19]. In our experiment with tubules, the surface velocity towards cathode being much smaller than the main flow towards anode, we conclude that the profile is not a pure electroosmotic one. Thus the tubules have an electrophoretic mobility indicating that they behave as negatively charged objects.

The values for tubules electrophoretic mobilities appeared as high as $3 \times 10^{-8}$ m$^2$ s$^{-1}$ V$^{-1}$. Such values are too high to be only due to charged impurities or to the hydrolysis of a few phosphate polar heads. However, this negative mobility can be due to the conformation of the lipid molecule in the chain frozen $L_{\beta'}$ phase. In this solid gel state with tilted hydrocarbon chains, the polar head is also tilted and the P-N axis is in the conformation shown on Fig. 16 [23]. In this case, the negative charge borne by the phosphate group can be more exposed to the solvent than the positive one on the choline group.

A negative apparent charge is also in good accordance with the behavior of the tubules with anionic or cationic particles. When tubules are dispersed with cationic magnetic particles, electron microscopy pictures shows that most tubules are full of particles, and magnetophoresis experiments confirm this result, as they show that tubules behave as objects more magnetic than the external medium (the magnetic fluid). On the contrary, when tubules are dispersed with anionic magnetic particles, these latter do not seem to enter inside the lumen of the tubules, and in the magnetophoretic experiments, tubules behave as objects less magnetic than the external medium.

These results point out the electrostatic interactions between tubules and particles, according to the sign of surface charges of the latter. When particles are positively charged, they are strongly attracted by the tubules, as if the phospholipids self-assemblies were negatively charged objects.

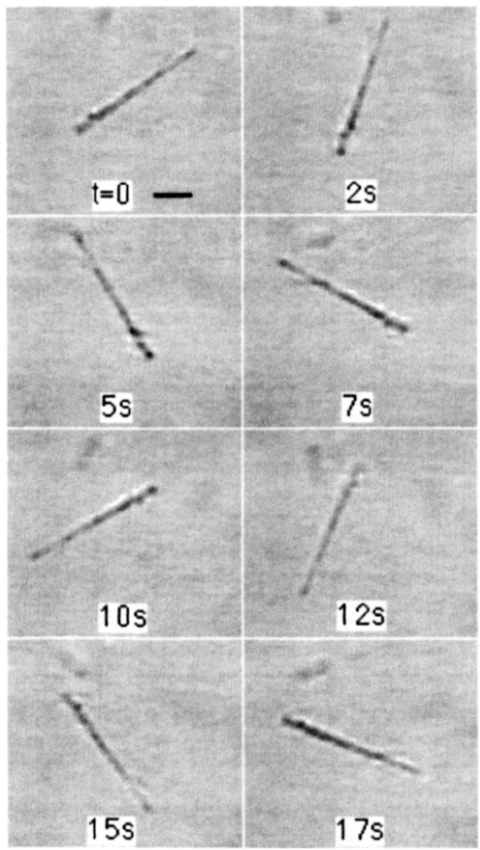

Fig. 15. Synchronous rotation of a tubule in a cationic magnetic fluid ($\phi$ = 2%) submitted to a rotating magnetic field (intensity $6.3 \times 10^3$ A m$^{-1}$ (80 G), frequency 0.05 Hz). The period T is 20 s and the time increment between two consecutive pictures of the same column is T/4=5s. The bar length is 5 µm.

### 3.2. Behavior of tubules in a magnetic fluid: "susceptibility" of tubules

The relevant parameter to describe the orientation of a tubule in a magnetic field is the relative difference between the magnetic permeability of the tubule ($\mu_{tub}$) and the one of the magnetic fluid ($\mu_{FF}$). We shall define the ratio $\chi$ (which is not a real susceptibility) as:

$$\chi = \frac{\mu_{tub} - \mu_{FF}}{\mu_{FF}} \qquad (1)$$

$\chi$ is only a convenient ratio which sign indicates if the tubule is more or less magnetic than the surrounding medium.

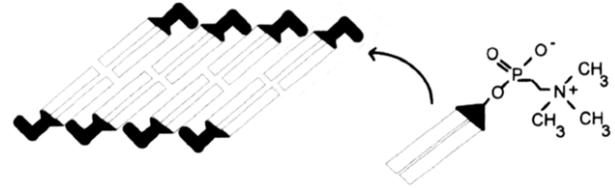

Fig. 16. Phospholipidic bilayer in the gel state ($L_{\beta'}$). The polar head is tilted.

### 3.3. Magnetophoresis: effect of a magnetic field gradient

Magnetophoresis is sensitive to the sign of the $\chi$ parameter. The field $H_x$ orients the tubule along the Ox axis while the field gradient causes its migration along Oy with a constant velocity $\vec{v}$. The tubule is described as a cylinder of radius r and length l. Its motion is driven by the magnetic force $\vec{F}_m$:

$$\vec{F}_m = \chi \mu_{FF} V H_x \left( \vec{\nabla}_y H_x \right) \qquad (2)$$

In fact we use dilute magnetic fluids with a magnetic susceptibility smaller than $10^{-3}$, so that $\mu_{FF}$ can be approximated by $\mu_0$. Viscous friction on the tubule surface results in a drag force $\vec{F}_v$ given by (from [24]):

$$\vec{F}_v = 4\pi\eta \vec{v} \, l \Big/ \ln\left(\frac{4\eta}{\rho r v}\right) \qquad (3)$$

where $\eta$ and $\rho$ are respectively the viscosity and the volumic mass of the surrounding liquid. Because the volume fraction of the magnetic colloidal particles in the magnetic fluid is of a few percents, those values may be taken the same as for pure water ($\eta = 10^{-3}$ Pa s and $\rho = 10^3$ kg m$^{-3}$).

Both forces being proportional to l, their balance leads to a velocity $\vec{v}$ which does not depend on the tubule length:

$$\vec{v} = r^2 \ln\left(\frac{4\eta}{\rho r v}\right) \chi \mu_0 H_x \left( \vec{\nabla}_y H_x \right) \Big/ 4\eta \qquad (4)$$

The geometry of a field gradient (along Oy) perpendicular to the field direction (Ox) enables to sort the tubules by their $\chi$ value but not by their size, as it is observed experimentally.

Under a field $H_x = 2.4 \times 10^4$ A m$^{-1}$ (300 G) and a field gradient $9.6 \times 10^5$ A m$^{-2}$ (120 G cm$^{-1}$), the mean velocity is $-15$ µm s$^{-1}$ in the anionic magnetic fluid ($\phi$ = 1.8%) and $+0.9$ µm s$^{-1}$ in the cationic magnetic fluid ($\phi$ = 2.0%). A typical value for r measured on the electronic micrographs is 0.25 µm. Thus equation 4 gives $\chi = -0.20$ for the anionic magnetic fluid and $\chi = +0.12$ for the cationic one.

### 3.4. Effect of a rotating magnetic field

When a rotating magnetic field $\vec{H}$ is applied to tubules, their long axes (Fig. 17) are not parallel to $\vec{H}$, and they bear a magnetic torque $\Gamma_m$. It is exactly balanced by a viscous torque $\Gamma_v$ in the stationary rotation regime. In this part we have to consider the tubules as elongated ellipsoids for the viscous torque on an ellipsoid rotating with angular velocity $\Omega$ in a liquid of viscosity $\eta$ has a known expression proposed by Perrin for a>5b [25]:

$$\Gamma_v = 4\eta V \frac{\left(\frac{a}{b}\right)^2}{2\ln\left(\frac{2a}{b}\right) - 1} \Omega \qquad (5)$$

The torque acting on an axisymetric ellipsoid of long semi-axis a, short semi-axis b, and volume V is [26]:

$$\Gamma_m = \frac{\chi^2 \mu_{FF}}{2(2+\chi)} H^2 V \sin(2\alpha) \qquad (6)$$

Stationary rotation is possible as long as there exists an angle $\alpha$ wich satisfies $\Gamma_m = \Gamma_V$. But when the angular velocity exceeds a critical value $\Omega_c$, the magnetic torque can no longer balance the visous one (the balance gives $\sin(2\alpha) > 1$). At the critical point $\sin(2\alpha) = 1$ and hence $\alpha = 45°$:

$$\Omega_c = \frac{\chi^2}{2+\chi} \mu_0 H^2 \frac{2\ln\left(\frac{2a}{b}\right) - 1}{8\eta\left(\frac{a}{b}\right)^2} \qquad (7)$$

In the case of a tubules in an anionic ferrofluid (volume fraction of particles 3.5%) we measured $\Omega_c/2\pi = 0.5$Hz for $H = 4.0 \times 10^3$ A m$^{-1}$ (51 Gauss). Assuming an aspect ratio $a/b = 10$ (Fig. 12) this second order equation gives a negative root, $\chi = -0.21$. In the other case, i.e. tubules in a cationic magnetic fluid ($\phi = 1.9\%$), $\Omega_c/2\pi = 0.05$Hz for $H = 6.3 \times 10^3$ A m$^{-1}$ (80 Gauss) and $a/b = 20$ (Fig. 15), we find a positive root $\chi = +0.08$. The choice between the roots of opposite signs in the precedent estimations is driven by the magnetophoresis experiment, which is sensible to the sign of the $\chi$ parameter.

Even if the so calculated values of the $\chi$ parameters are only approximative, the three experiments on tubules in a magnetic fluid (alignment by a static filed, stationary rotation and magnetophoresis) lead to the same result: the tubules are less magnetic than the carrier liquid ($\chi < 0$) when this one is an anionic magnetic fluid, although they are more magnetic than the surrounding medium ($\chi > 0$) when the magnetic particles are cationic. Besides, the lack of magnetic permeability in the fisrt case is higher than the excess in the second case. This result accounts for the following conclusion: in an anionic magnetic fluid, tubules and negatively charged particles do not interact, and tubules behave as magnetic holes [27] in a magnetic medium; in a cationic magnetic fluid, particles stick on the tubules even fill them, and tubules become more magnetic than the medium. Nevertheless, in this last case, the phospholipidic membrane which is not magnetic lowers the $\chi$ value.

## 4. Conclusion and prospects

The use of molecular self-assembly of lipids for the fabrication of micrometric magnetic tubules is reported here. Two kinds of aqueous magnetic fluid (with cationic or anionic particles) lead to different systems and well-separated behaviors. The characterization of the magnetic tubules reveals an electrostatic interaction between the tubules behaving as negative objects and the particles according to the sign of their surface charges. Electron microscopy allows to interpret these experiments: anionic particles do not interact with tubules surface, whereas cationic particles stick to the membrane and seem to fill the tubules. It appeared that the tubules dispersed in a magnetic fluid in every case orientate along the direction of an applied magnetic field of low intensity and follow a rotating magnetic field. In a magnetic field gradient, the way the tubules moves is function of the sign of particles surface charges. Tubules full of cationic particles are more magnetic than the surrounding medium and tubules mixed with anionic particles are less magnetic.

These results illustrate again the possibility to use magnetic fluids either as an "inert" medium for the orientation of biological species or as agents able to modify self-assemblies of phospholipids in order to obtain magnetic hybrid organic–inorganic objects.

## Acknowledgements

The authors thank D. Riveline for his help with the electrophoresis experiments, J. Servais for the engineering of the electronic setups and J-C. Bacri for his comments on the manuscript.